\documentclass[12pt]{article}
\usepackage{amsmath}
\usepackage{graphicx,psfrag,epsf}
\usepackage{enumerate}
\usepackage{natbib}
\usepackage{url} 

\usepackage{enumerate}
\usepackage{appendix}
\usepackage{enumitem}
\usepackage{booktabs}
\usepackage{adjustbox}
\usepackage{subcaption}
\usepackage{amssymb}

\newtheorem{assumption}{Assumption}
\newtheorem{proposition}{Proposition}

\def\bSig\mathbf{\Sigma}

\newcommand{\Yobs}{Y_i^{obs}}
\newcommand{\Ynmi}{Y_i^{nmi}}
\newcommand{\Ymi}{Y_i^{mi}}
\newcommand{\Ymistar}{Y_i^{mi, *}}

\newcommand{\Dobs}{D_i^{obs}}
\newcommand{\Dcur}{D_i^{cur}}
\newcommand{\Dno}{D_i^{no.nmi}}

\newcommand{\Tnmi}{T_i^{nmi}}
\newcommand{\Tmi}{T_i^{mi}}

\newcommand{\blind}{0}

\date{}
\begin{document}

\def\spacingset#1{\renewcommand{\baselinestretch}%
{#1}\small\normalsize} \spacingset{1}


\if0\blind
{
  \title{\bf Estimating the Malaria Attributable Fever Fraction Accounting for Parasites Being Killed by Fever and Measurement Error}
  \author{Kwonsang Lee\thanks{Kwonsang Lee is Ph.D. student, Department of Statistics and Applied Mathematics and Computer Science Program, University of Pennsylvania, Philadelphia, PA 19104 (E-mail: kwonlee@wharton.upenn.edu)} \,\,and Dylan S. Small\thanks{Dylan S. Small is Professor, Department of Statistics, The Wharton School, University of Pennsylvania, Philadelphia, PA 19104 (E-mail: dsmall@wharton.upenn.edu). The Authors thank Thomas Smith for insightful suggestions and providing the data.}
   \hspace{.2cm}\\   
}
  \maketitle
} \fi

\if1\blind
{
  \bigskip
  \bigskip
  \bigskip
  \begin{center}
    {\large\bf Estimating the Malaria Attributable Fever Fraction Accounting for Fever Killing Parasites Being Killed by Fever and Measurement Error}
\end{center}
  \medskip
} \fi

\bigskip
\begin{abstract}
Malaria is a parasitic disease that is a major health problem in many tropical regions. The most characteristic symptom of malaria is fever. The fraction of fevers that are attributable to malaria, the malaria attributable fever fraction (MAFF), is an important public health measure for assessing the effect of malaria control programs and other purposes. Estimating the MAFF is not straightforward because there is no gold standard diagnosis of a malaria attributable fever; an individual can have malaria parasites in her blood and a fever, but the individual may have developed partial immunity that allows her to tolerate the parasites and the fever is being caused by another infection. We define the MAFF using the potential outcome framework for causal inference and show what assumptions underlie current estimation methods. Current estimation methods rely on an assumption that the parasite density is correctly measured. However, this assumption does not generally hold because (i) fever kills some parasites and (ii) the measurement of parasite density has measurement error. In the presence of these problems, we show current estimation methods do not perform well. We propose a novel maximum likelihood estimation method based on exponential family g-modeling. Under the assumption that the measurement error mechanism and the magnitude of the fever killing effect are known, we show that our proposed method provides approximately unbiased estimates of the MAFF in simulation studies. A sensitivity analysis can be used to assess the impact of different magnitudes of fever killing and different measurement error mechanisms. We apply our proposed method to estimate the MAFF in Kilombero, Tanzania. 
\end{abstract}

\noindent%
\vfill
\newpage
\spacingset{1.45} 
\section{INTRODUCTION}
\label{s:intro}

Malaria is a mosquito-borne infectious disease caused by a parasite. In many tropical regions, malaria is a giant-killer of children, imposes financial hardship on poor households, and holds back economic growth and improvements in living standards (World Health Organization, 2011). The most characteristic clinical feature of malaria is fever \citep{warrell1993clinical}.  The malaria attributable fever fraction (MAFF) for a group of people is the proportion of fevers in the group of people that are attributable to (caused by) malaria. We will consider the MAFF for children in sub-Saharan Africa, the population hit hardest by malaria (World Health Organization, 2011).  The MAFF is an important public health quantity for several reasons that include
\begin{itemize}
\item The MAFF provides information about the public health burden from malaria and how much resources should be devoted to combatting malaria compared to other diseases \citep{mabunda2009country}.

\item The MAFF is an essential input to the prevalence of malaria attributable fevers (PMAF); the PMAF equals MAFF $\times$ prevalence of fevers. Changes in the PMAF over time provide information about the effects of public health programs that combat malaria \citep{koram2007malaria}. 

\item For planning the sample size for a clinical trial of an intervention against malaria, the MAFF is an essential input \citep{halloran1999design, smith2007measures}.  For example, suppose we are planning a trial of duration one year and want to have 80\% power for an intervention that halves malaria attributable fever but has no effect on other sources of fever, and in the population of interest, the average child suffers from 10 fevers per year.  The needed sample size depends on the MAFF and is about $n=(800, 1325, 2500)$ in this example if the MAFF is $(0.5, 0.4, 0.3)$ respectively. The details of the sample size calculation are in the Supplementary Materials.  

\item For clinicians treating a child suffering from fever and needing to decide how to prioritize providing antimalarial treatment vs. treatments for other possible sources of the fever, knowing the MAFF conditional on the child's symptoms (e.g., the intensity of the fever and the child's  parasite density) is a valuable input \citep{koram2007malaria}. In particular, letting `MAFF $|$ Symptoms' denote the MAFF conditional on the child's symptoms, a doctor would want to treat a patient with an anti-malarial if (Expected gain in utility from treating child at time $t$ with an anti-malarial if she has a malaria attributable fever vs. not treating)$\times$(MAFF $|$ Symptoms) $>$  (Expected loss in utility from treating child at time $t$ with an anti-malarial if she has a fever that is not malaria attributable fever vs. not treating) $\times$ [1-(MAFF $|$ Symptoms)].

\end{itemize}

The MAFF could be estimated from a survey by a usual ratio estimator if it was easy to determine whether or not a fever was attributable to malaria parasites.  However, fevers caused by malaria parasites often cannot be distinguished on the basis of clinical features from fevers caused by other common childhood infections such as the common cold, pneumonia, influenza, viral hepatitis or typhoid fever \citep{hommel2002diagnostic, koram2007malaria}. One aid to deciding whether a fever is caused by malaria parasites or some other infection is to measure the density of malaria parasites in the child's blood.  However, in areas where malaria is highly endemic, children can develop partial immunity to the toxic effects of the parasites and can tolerate high parasite densities without developing fever \citep{marsh2002immunology,boutlis2006malaria}. Consequently, even if a child has a fever and has a high parasite density, the fever might still be caused by another infection.  In summary, it cannot be determined with certainty whether a given child's fever is malaria attributable, making estimating the MAFF challenging.

In this paper, we make two contributions to estimating the MAFF. First, we provide an analysis of the assumptions needed for existing estimators of the MAFF to be consistent and show that these assumptions are not plausible. In previous work on estimating the MAFF, estimators have been proposed without clearly defining the estimand. We use the potential outcomes framework to clearly define the estimand and make clear the causal assumptions on which the consistency of these previous estimators rest. These assumptions include that non-malaria attributable fevers do not kill parasites and that there is no measurement error of a certain type in measuring parasite density. We discuss evidence that these assumptions are {\it not} plausible in most settings, and show that existing MAFF estimation methods are biased under plausible violations of the assumptions. 

The second major contribution of our paper is that we develop a consistent estimator of the MAFF that allows for parasites being killed by fever and measurement error in parasite density. Our novel estimation method extends the g-modeling method for solving deconvolution problems \citep{efron2016empirical} to the setting of malaria survey data, accounting for parasites being killed by fever and measurement error. Specifically, survey data on malaria can be divided into two groups, the children with fever (the febrile group) and the children without fever (the afebrile group). The group with fever is a mixture of two components: children with a fever that is malaria attributable and children with a fever that is not malaria attributable. The group without fever can be used as a training sample that provides information about the distribution of parasite densities of the latter mixture component. The main idea of our method is to recover the distributions of the parasite densities of the mixture components before fever killing and measurement error by assuming that the mixture components are in flexible exponential families and solving the deconvolution problem. Using simulation studies, we show that our proposed method produces approximately unbiased estimates of the MAFF when the magnitude of fever killing and the measurement error mechanism are correctly specified. We apply our method to make inferences about the MAFF for a study area in Kilombero, Tanzania. 

The rest of this article is organized as follows. In Section~\ref{s:maff}, we define the MAFF and state critical assumptions in the potential outcome framework. We also reveal the relationship between the potential outcome framework MAFF and the probability limits of the conventionally used estimators of the MAFF based on observable quantities under the assumptions. Section~\ref{s:fever} describes possible violations of the assumptions that there is no fever killing and no measurement error. Also, the impact of a violation of these assumptions on current estimation methods is investigated. In Section~\ref{s:model}, we develop our new estimation method, the maximum likelihood estimation method using the g-modeling approach. Section~\ref{s:sim} shows the performance of our proposed method with simulation studies and Section~\ref{s:example} shows the application to the malaria data in Tanzania with a sensitivity analysis. Section~\ref{s:summary} includes summary and discussion. 

\section{MALARIA ATTRIBUTABLE FEVER FRACTION (MAFF)}
\label{s:maff}

\subsection{Potential Outcome Definition}
\label{ss:sec2_definition}

To define a fever as being caused by (i.e., attributable to) malaria parasites, we use the Neyman-Rubin potential outcomes framework for causal inference \citep{splawa1923proba, rubin1974estimating}. For each child $i$ and each possible parasite density level $d$, the potential outcome $Y_i^{(d, z)}$ is $1$ or $0$ according to whether or not the child would have fever if an intervention set the child's parasite density level to $d$ and the amount of non-malarial infections to $z$. (the intervention does not need to be specified but need to satisfy Assumption~\ref{assumption2} which we will discuss below). Each child has many potential outcomes, but we observe only one potential outcome, $Y_i^{obs}\equiv Y_i^{(D_i, Z_i)}$, where $D_i$ is the child's actual parasite density (we are only able to measure $D_i$ with some error; see Section~\ref{s:fever}) and $Z_i$ is the amount of non-malarial infections that the child has (we cannot directly observe $Z_i$, but assume that $Z_i=0$ means no non-malarial infection and higher levels of $Z_i$ mean more non-malarial infections). In addition to the actually observed outcome $Y_i^{obs}$, we also consider the potential outcome under an intervention that eliminates malaria parasites from the child's body, $\Ynmi \equiv Y_i^{(0, Z_i)}$; the `nmi' stands for whether the child would have a fever if the malaria parasites were eliminated and the only possible source of fever were a non-malarial infection (nmi). Let $T_i^{nmi}$ be the threshold of non-malaria infection $Z$ that would be needed to trigger a non-malarial fever if the malaria parasites were eliminated. If $Z_i$ is greater than or equal to $T_i^{nmi}$, then we define the child to have a non-malarial fever (i.e. $\Ynmi=1$).  If $\Ynmi=1$, then the child has a fever that is not attributable to malaria because the fever would be present even if the malaria parasites were eliminated. 

Malaria parasites are thought to cause fever by causing the red blood cells they invade to rupture, releasing an insoluble hemoglobin digestion product, hemozoin, which stimulates innate immune responses that cause the body's temperature to rise \citep{parroche2007malaria}. A child's pyrogenic threshold is the minimum malaria parasite density at which these immune responses will result in a fever if there is no non-malarial infection; the pyrogenic threshold varies from child to child based on factors such as the child's acquired immunity (through previous exposure to malaria) and natural immunity (e.g., genetic inheritance of the sickle cell trait) \citep{rogier1996evidence, langhorne2008immunity, taylor2012haemoglobinopathies}. Let $T_i^{mi}$ denote child $i$'s pyrogenic threshold. 
 We denote $\Ymi \equiv Y^{(D_i, 0)}$ as a variable that indicates whether the density exceeds the threshold $\Tmi$, i.e., whether the child's malaria infection (mi) is strong enough to cause a fever in the absence of any non-malaria infection; $\Ymi=1$ if $D_i \geq \Tmi$ and $\Ymi=0$ otherwise. This definition implies that if $D_i \geq \Tmi$, then the child has a malaria fever ($\Ymi=1$), thus resulting in $\Yobs =1$. However, the opposite is not true. Even if $\Yobs=1$, we do not know whether $\Ymi$ is 1 or not. It is possible that the fever is caused only by a non-malaria infection, $\Ynmi=1$ but $\Ymi=0$ (meaning $D_i < \Tmi$). Note that it is possible that both $\Ymi=1$ and $\Ynmi=1$, meaning that the child has both a malaria infection that exceeds the child's pyrogenic threshold and a non-malarial infection that would cause a fever even if the child's malaria parasites were eliminated. We assume for now that the way malaria and non-malaria infections affect a fever is like parallel circuits so that a fever happens if and only if $Y^{mi}=1$ and/or $Y^{nmi}=1$, i.e., the observed outcome $\Yobs$ can be represented by a function of $\Ynmi$ and $\Ymi$: $\Yobs= \min \{ \Ynmi+\Ymi, 1 \}$. This is specified as Assumption~\ref{assumption1} (iv) below. We will relax this assumption in Section~\ref{ss:violation}.

We define a child $i$ who is observed to have a fever as having a {\it{malaria attributable fever}} if that fever would not have occurred if the child had been given an intervention that prevented the child from having malaria parasites. In terms of potential outcomes, a child $i$ has a malaria attributable fever if $Y_i^{obs}=1$ but $Y_i^{nmi}=0$. Let $(Y^{mi},Y^{nmi})$ be the random vector from the experiment of choosing a random child and time point from the study area and study period.  The fraction of fevers in the study area and time period that are attributable to malaria, i.e., the MAFF, is
\begin{equation}
MAFF=P(Y^{nmi}=0,Y^{obs}=1|Y^{obs}=1).
\label{maff.defn}
\end{equation}

Because we have defined the MAFF using potential outcomes based on an intervention that could alter parasite density, the MAFF could depend on the intervention that alters parasite density, e.g., possible interventions are antimalarial drugs, vaccines and control of the mosquitoes that carry the malaria parasites. Consider an intervention that satisfies:
\begin{assumption}
\normalfont {\it{No Side Effects Assumption}}.  The intervention has no effects on fever beyond removing
the parasites from the child.
\label{assumption2}
\end{assumption}
Assumption~\ref{assumption2} implies that the intervention cannot cause a fever:
\begin{equation}
P(Y^{nmi}=1,Y^{obs}=0)=0. \label{eq:implication.A1}
\end{equation}
Under the no side effects assumption, the MAFF can be interpreted as the proportion of fevers in the observed world that would be eliminated if malaria were eradicated by the intervention.  We will not specify the intervention under consideration but will assume that the intervention satisfies Assumption~\ref{assumption2}.  Although it is possible that interventions currently being studied could have side effects and violate Assumption~\ref{assumption2}, the MAFF for a hypothetical intervention that satisfies Assumption~\ref{assumption2} provides an estimate of the {\it{potential}} benefit of an intervention to eliminate malaria which is useful for policymakers \citep{walter1976estimation}.

The potential outcome framework MAFF~(\ref{maff.defn}) is defined based on unobservable variables $Y^{nmi}$ and $Y^{mi}$. In order to identify the MAFF from data, assumptions need to be made about the way these unobserved variables link to observed variables. In the following section, we discuss identifying assumptions.

\subsection{Assumption}
\label{ss:sec2_assumption}

To estimate the MAFF, we make the following assumptions: 
\begin{assumption}
\normalfont The potential outcome $Y^{(d, z)}$ and the pyrogenic thresholds $\Tnmi$ and $\Tmi$ satisfy that
\begin{itemize}
\item[(i)] (Monotonicity Assumption) 
\begin{itemize}
\item if $d \leq d'$, then $Y^{(d, z)} \leq Y^{(d', z)}$ for all $z$.

\item if $z \leq z'$, then $Y^{(d, z)} \leq Y^{(d, z')}$ for all $d$.
\end{itemize}

\item[(ii)] the pyrogenic thresholds $T^{nmi}$ and $T^{mi}$ are strictly greater than zero. 

\item[(iii)] a fever caused by a non-malaria infection $Y^{nmi}$ is independent of a fever caused by a malaria infection $Y^{mi}$, i.e. $Y^{nmi} \perp\!\!\!\perp Y^{mi}$.

\item[(iv)] $Y^{obs}= \min \{Y^{nmi}+Y^{mi}, 1 \}$

\end{itemize}
\label{assumption1}
\end{assumption}
Assumption~\ref{assumption1} (i), the monotonicity assumption, is biologically plausible because having more parasites means that more red blood cells are ruptured by the parasites and more hemozoin is released, meaning that if a child's hemozoin level was enough to cause a fever at parasite level $d$, then a child would surely have a fever at parasite level $d'$ since the child's hemozoin level would be even higher. Similarly, the monotonicity assumption for non-malaria infections $z$ is plausible. Assumption~\ref{assumption1} (ii) is also biologically plausible because it says that a fever cannot be triggered by malaria parasites if there are no malaria parasites present. Similarly, it is also plausible for non-malarial infections. Assumption~\ref{assumption1} implicitly defines $Y^{(T^{mi}, Z)}=1$ for the pyrogenic threshold $T^{mi}$. If the parasite density reaches the pyrogenic threshold, then a fever is observed. Furthermore, it follows that $Y^{(d, z)}=1$ for any $d > T^{mi}$. We emphasize that this assumption does not imply that $Y^{(d, z)}=0$ if $d < \Tmi$; the child could have a fever from a non-malarial infection, $Y^{nmi}=1$. Assumption~\ref{assumption1} (iii) means that the potential outcome $Y^{nmi}$ is independent of the indicator $I(D \geq T^{mi})$. This assumption implies that there is no causal pathway between the parasite density $D$ and $Y^{nmi}$, and also no causal pathway between the pyrogenic threshold $T^{mi}$ and $Y^{nmi}$. This assumption may be more plausible after controlling for observed covariates and we show how we  control for observed covariates in Section~\ref{ss:violation}. Also, the impact of the violation of the assumption is investigated by conducting a sensitivity analysis. Assumption~\ref{assumption1} (iv) is the assumption discussed in Section~\ref{ss:sec2_definition} that the way malaria and non-malaria infections affect a fever is like parallel circuits. We will develop a sensitivity analysis for violations of this assumption in Section~\ref{ss:violation}. 

\subsection{Additional Assumptions Needed for Existing Estimators}
\label{ss:sec2_properties}

We will show in this section that the existing estimators of the MAFF are not consistent under Assumptions~\ref{assumption2} and \ref{assumption1} alone, and that an additional, implausible assumption is needed. Let $p_f$ be the parasite prevalence in febrile children and $p_a$ be the parasite prevalence in afebrile children. That is, $p_f= P(D^{obs}>0 | Y^{obs}=1)$ and $p_a= P(D^{obs}>0 | Y^{obs}=0)$ where $D^{obs}$ is the observed parasite density. One popular estimator is based on the relative risk \citep{smith1994attributable}, which is equal to $\hat{p}_f (\hat{R}-1)/\hat{R}$ where $\hat{p_f} = \hat{P}(D^{obs}>0 | Y^{obs}=1)$ and $\hat{R} = \hat{P}(Y^{obs}=1|D^{obs}>0)/\hat{P}(Y^{obs}=1|D^{obs}=0)$. This estimator is proposed as the population attributable fraction estimator by \citet{greenwood1987mortality}, and discussed in \citet{smith1994attributable}. The estimator of the MAFF based on relative risk converges in probability to
\begin{equation}
\mbox{plim}(\widehat{MAFF}_{RR})  = p_f (R-1)/R
\label{def:maff}
\end{equation}
where $R$ is the relative risk of fever associated with the exposure of parasites, i.e. $R = P(Y^{obs}=1 | D^{obs}>0) / P(Y^{obs}=1 | D^{obs}=0)$. We note that $p_a, p_f$ and $R$ can be estimated from the observed data $(Y^{obs}, D^{obs})$. The consistency of the estimator $\widehat{MAFF}_{RR}$ relies on the following assumption. 
\begin{assumption}
\normalfont {\it No Errors Assumption.} The parasite density does not change before and after having a fever caused solely by a non-malaria infection, $Y^{nmi}=1, Y^{mi}=0$, and the observed parasite density $D^{obs}$ is measured without error.
\label{assumption3}
\end{assumption}
In Section~\ref{s:fever}, we will discuss that the no errors assumption is implausible. Under Assumptions \ref{assumption2} - \ref{assumption1} and the no errors assumption, the following proposition reveals that $\widehat{MAFF}_{RR}$ is consistent.

\begin{proposition}
\normalfont Under Assumptions~\ref{assumption2} - \ref{assumption3}, the potential outcome framework MAFF is equal to the relative risk MAFF. That is, $MAFF=\mbox{plim}(\widehat{MAFF}_{RR})$.
\label{proposition1}
\end{proposition}
The proof is provided in the Appendix. 

Another popular choice of an estimator for the MAFF based on odds ratio is $\widehat{MAFF}_{OR}=({\hat{p}_f - \hat{p}_a})/({1-\hat{p}_a})$ that has the probability limit, $\mbox{plim}(\widehat{MAFF}_{OR})$. This estimator is an approximated version of the estimator $\hat{p}_f (\hat{R}-1)/\hat{R}$ because the relative risk $R$ is often approximated by the odds ratio, $p_f(1-p_a)/p_a(1-p_f)$. Then, the probability limit of this estimator is obtained as 
\begin{equation}
\mbox{plim}(\widehat{MAFF}_{OR}) = (p_f-p_a)/(1-p_a).
\label{def:maff_adj}
\end{equation}
$\mbox{plim}(\widehat{MAFF}_{OR})$ is approximately equal to $\mbox{plim}(\widehat{MAFF}_{RR})$ when the prevalence of cases is rare. 
\begin{proposition}
\normalfont Under Assumptions~\ref{assumption2}-\ref{assumption3}, in terms of the potential outcome framework, $\mbox{plim}(\widehat{MAFF}_{OR})$ is given by
\begin{equation}
\mbox{plim}(\widehat{MAFF}_{OR})= \frac{P(Y^{mi}=1)}{P(Y^{obs}=1)} = \frac{\mbox{plim}(\widehat{MAFF}_{RR})}{P(Y^{nmi}=0)}.
\label{eqn:maff_pot2}
\end{equation}
\label{proposition2}
\end{proposition}
The proof is also provided in the Appendix. According to Proposition~\ref{proposition2}, under Assumptions~\ref{assumption1} and \ref{assumption3}, $\widehat{MAFF}_{OR}$ is an asymptotically biased estimator of the MAFF, and estimates the proportion of children who have malaria fevers among febrile children $P(Y^{mi}=1 | Y^{obs}=1)$, not the MAFF.  

 Proposition~\ref{proposition2} implies that $\mbox{plim}(\widehat{MAFF}_{OR})$ is strictly larger than $\mbox{plim}(\widehat{MAFF}_{RR})$ when the probability of having non-malaria caused fever is positive. Technically, $\mbox{plim}(\widehat{MAFF}_{RR})$ is represented by multiplication of the probability of not having non-malaria caused fever $P(Y^{nmi}=0)$ and $\mbox{plim}(\widehat{MAFF}_{OR})$ as shown by equation~(\ref{eqn:maff_pot2}). If the target estimand of some methods is $\mbox{plim}(\widehat{MAFF}_{OR})$, the estimate from the method should be adjusted by multiplying $P(Y^{nmi}=0)$ in order to acquire the estimate of $\mbox{plim}(\widehat{MAFF}_{RR})$. However, one difficulty is that $P(Y^{nmi}=0)$ is not observable. The following proposition shows that this adjustment can be successfully achieved by estimating $P(Y^{nmi}=0)$. 
\begin{proposition}
\normalfont Under Assumption~\ref{assumption2}-\ref{assumption3}, the estimator $\widehat{MAFF}_{RR}$ can be represented by the estimator $\widehat{MAFF}_{OR}$ and the probability ${p}=\widehat{P(Y^{obs}=1)}$. Let $\lambda=\widehat{MAFF}_{RR}$ and $\lambda^*=\widehat{MAFF}_{OR}$. Then, $\lambda$ is obtained as
\begin{equation}
\lambda = \frac{\lambda^*-p\lambda^*}{1-p\lambda^*}.
\label{eqn:maff_relation}
\end{equation} 
\label{proposition3}
\end{proposition}

{Proof. 
From equation~(\ref{eqn:maff_pot2}),
\begin{eqnarray*}
\lambda &=& P(Y^{nmi}=0) \cdot \lambda^* \\
&=& \frac{P(Y^{mi}=0)P(Y^{nmi}=0)}{P(Y^{mi}=0)} \cdot \lambda^* \\
&=& \frac{1-P(Y^{obs}=1)}{1-P(Y^{mi}=1)} \cdot \lambda^* \\
&=& \frac{1-p}{1-p\lambda^*} \cdot \lambda^* .
\end{eqnarray*}
}

There have been many estimators developed to estimate the MAFF. These estimators mainly can be classified as belonging to two different approaches through either estimating $\mbox{plim}(\widehat{MAFF}_{RR})$ or estimating $\mbox{plim}(\widehat{MAFF}_{OR})$. The first approach is to estimate the probability of having a fever given a parasite density $D$, $P(Y^{obs}=1|D)$, in order to estimate $\mbox{plim}(\widehat{MAFF}_{RR})$. This estimate leads to an estimate of the relative risk $R$ to obtain an estimator $\widehat{MAFF}_{RR}$. The simplest model is using the logistic regression to estimate the probability $P(Y^{obs}=1|D)$ \citep{smith1994attributable}. A slightly more complicated model that was preferred by \citet{smith1994attributable} adds an additional power parameter to the logistic regression. Also, nonparametric regression models can be considered such as local linear smoothing \citep{wang2012comparative}. However, if data are collected from a case-control study and the prevalence of fever is not known, then $P(Y^{obs}=1 | D)$ cannot be fully estimated, but $p_f$ and $p_a$ can be estimated. In this situation, the second approach is to estimate the probability density functions in both the febrile and afebrile subgroups (i.e. estimate the densities $f(D|Y^{obs}=1)$ and $f(D|Y^{obs}=0)$) in order to estimate $\mbox{plim}(\widehat{MAFF}_{OR})$. The febrile subgroup consists of samples from both the non-malaria caused fever population and the malaria caused fever population. The afebrile subgroup is a sample from the non-malaria caused fever population in the febrile subgroup. Under the no errors assumption, the parasite density of the non-malaria caused fever population has the same distribution as the parasite density of the afebrile subgroup. The afebrile subgroup is used as training data to deconvolve the mixture in the febrile subgroup. This data structure is sometimes called a two-component mixture problem. To tackle the two component mixture problem, semiparametric estimation models using empirical likelihood \citep{qin2005semiparametric} and Bayesian estimation \citep{vounatsou1998bayesian} have been considered. These models provide a consistent estimate of $plim(\widehat{MAFF}_{OR})$. However, $plim(\widehat{MAFF}_{OR})$ does not equal the MAFF even under the no errors assumption - see Proposition~\ref{proposition2}. Therefore, an additional step is required to adjust the obtained estimator $\widehat{MAFF}_{OR}$ using equation~(\ref{eqn:maff_relation}) to obtain an estimator that would be consistent under Assumptions~\ref{assumption2}-\ref{assumption3}. 

The new estimation method we will develop in Section~\ref{s:model} is based on estimating $\mbox{plim}(\widehat{MAFF}_{OR})$ and then adjusting the estimate by multiplying it by $\widehat{P(Y^{nmi}=0)}$ to estimate the MAFF. Using $\widehat{MAFF}_{RR}$ is more straightforward than using $\widehat{MAFF}_{OR}$ because it does not require the adjustment. However, the estimator $\widehat{MAFF}_{RR}$ can perform poorly when the proportion of afebrile children without parasites is small. The performance of $\widehat{MAFF}_{RR}$ depends on the precision of estimating $P(Y^{obs}=1|D)$, but the estimation can be noisy in usual malaria data. More importantly, the approach that uses the estimator $\widehat{MAFF}_{OR}$ considers a two-component mixture data structure. This mixture structure facilitates accounting for the fever killing effect and measurement error that will be discussed in the following section. 

\section{FEVER KILLING EFFECT AND MEASUREMENT ERROR}
\label{s:fever}

Under Assumptions~\ref{assumption2}-\ref{assumption3}, we can estimate the MAFF by using observable quantities through computing $\widehat{MAFF}_{RR}$ (or $\widehat{MAFF}_{OR}$) from the discussed approaches in Section~\ref{ss:sec2_properties}. However, Assumption~\ref{assumption3} (the no errors assumption) is generally implausible becasue of fever killing and measurement error. Fever killing refers to the fact that a fever kills some parasites in the body and measurement error refers to the fact that it is difficult to measure parasite density with great accuracy. In this section, we provide a model describing fever killing and measurement error, and show the effect of these problems on the performance of the existing methods. 

To account for fever killing and measurement error, we consider three different variables related to the parasite density: $\Dno, \Dcur$ and $\Dobs$. Let $\Dno$ be the parasite density that a subject $i$ would have if the subject does not have a non-malaria infection strong enough to cause a fever \citep{small2010evaluating}, and let $\Dcur$ be the amount of the parasite density in blood of the subject after fever killing occurs. Also, let $\Dobs$ be the observed parasite density. If a subject $i$ has a fever that is solely caused by a non-malaria infection, $Y_i^{nmi}=1, Y_i^{mi}=0$, then there is evidence that fever kills some of the parasites that would have remained alive in the absence of the infection \citep{kwiatkowski1989febrile, rooth1992suppression, long2001plasmodium}. In particular, \citet{long2001plasmodium} estimate that a fever of 38.8$^{\circ}$C kills 50 \% of parasites and a fever of 40$^{\circ}$C kills 92 \% of parasites. Fever killing will make $D_i^{no.nmi}$ greater than the actual current parasite density. We note that $D_i^{no.nmi}=D_i^{cur}$ unless $Y_i^{nmi}=1, Y_i^{mi}=0$. See \citet{small2010evaluating} for more discussion on fever killing. 

Fever killing occurs in some sense for all fevers, however we define $\Dno$ for malarial fevers ($\Ymi=1$) in such a way that fever killing in terms of $\Dcur$ being different from $\Dno$ occurs only for non-malarial fevers $Y_i^{nmi}=1, Y_i^{mi}=0$ Specifically, we define $\Dno$ in the following way for a child with a malarial fever, $\Ymi=1$. In a malaria infection that is not brought under control by a child's immunity, the parasites multiply inside the red blood cells they invade, eventually causing the red blood cell to rupture, and the released parasites then invade new red blood cells \citep{kitchen1949falciparum}. This causes an exponential growth phase of the parasites that terminates shortly after the onset of periodic fever \citep{kitchen1949falciparum}. The fever starts killing parasites while at the same time the parasites that remain alive continue to multiply. The clash of these two forces and the features of the parasite life cycle cause the parasite density to oscillate \citep{kwiatkowski1991periodic}. For a child with a malaria fever $\Ymi=1$, even if any non-malarial infection were removed, this process of the parasites multiplying and eventually rising above the pyrogenic threshold and then oscillating would occur. For a child with a malarial fever, we define the child's parasite density $\Dno=\Dcur$ as the parasite density at the point in the child's fever at which the child is observed. 
 
  
Besides fever killing, measurement error occurs when researchers measure the actual current parasite density $\Dcur$. Let $\Dobs$ be the observed (measured) parasite density from a blood sample. \citet{dowling1966comparative} and \citet{o2007malaria} study the sources and magnitude of measurement error. The sources include the following: (1) Sample variability. The parasite density is estimated from a sample of blood; (2) Loss of parasites in the sample handling and staining process; (3) Microscopy error. The accuracy of the parasite density measurement depends on the quality of the microscope and the concentration and motivation of the microscopist; (4) Sequestration and synchronization. Microscopic examination of a blood sample only estimates the parasite density in the peripheral blood and not the total parasite density. Older parasites sequester in the vascular beds of organs. Due to a tendency of the life cycles of the parasites to be synchronized, there can be large variation in the parasite density in the peripheral blood relative to the total parasite density \citep{bouvier1997seasonality}; (5) Variability in white blood cell density.  The most common method of estimating parasite density counts the number of parasites found for a fixed number of white blood cells and then assumes that there are 8000 white blood cells per $\mu\mathrm{l}$ of blood.  White blood cell counts actually vary considerably from person to person and from time to time within a person \citep{mckenzie2005white}.

\begin{figure}[t]
\centerline{
\includegraphics[width=120mm]{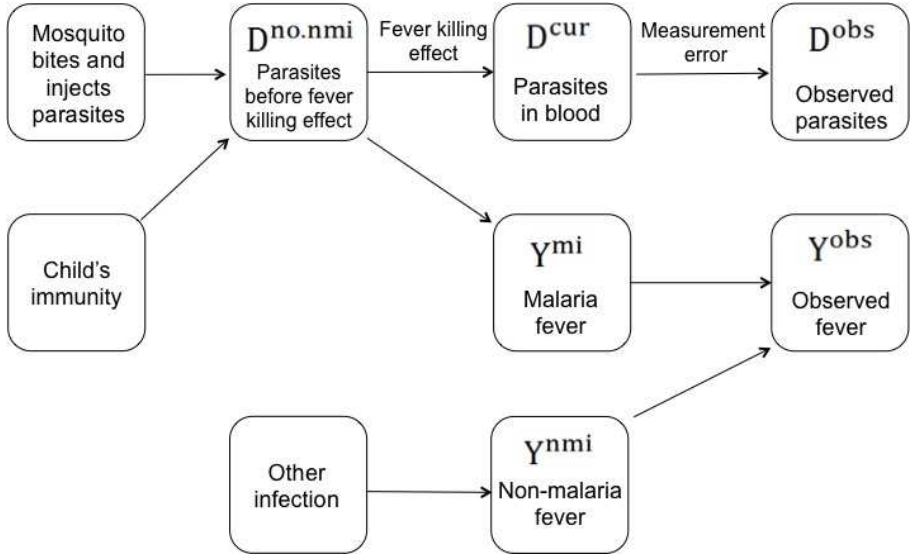}}
\caption{Causal diagram}
\label{fig:variables}
\end{figure}

Figure 1 summarizes the relationship between all the defined variables. As can be seen, $\Dno$ is the only parasite density variable that decides whether child $i$ has a malaria fever or not while $\Dcur$ and $\Dobs$ are proxy variables of $\Dno$. We can think of $\Dno$ as the parasitological  challenge faced by the child which is a function of the amount of the parasites injected from mosquito bites and the immune response of the child. The other two parasite density variables, $\Dcur$ and $\Dobs$, change according to both fever killing and measurement error but do not directly affect whether the child has a malarial fever.

\subsection{Simulation Study of Existing Methods}
\label{ss:simul_methods}

We evaluate the performance of several existing methods for estimating the MAFF with a simulation study. Specifically, we consider three settings; (1) there is no fever killing and no measurement error (we call it Situation 1), (2) there is no fever killing effect, but there is measurement error (Situation 2) and (3) there is both fever killing (50\%) and measurement error (Situation 3). We consider the four estimation methods discussed in Section~\ref{s:maff}: logistic regression (L), logistic regression with power parameter (P), local linear smoothing followed by isotonic regression (LI) and the adjusted semiparametric method (S). The first three methods (L, P and LI) have probability limits of $\mbox{plim}(\widehat{MAFF}_{RR})$. The semiparametric model method has a probability limit of $\mbox{plim}(\widehat{MAFF}_{OR})$, so we use adjustment~(\ref{eqn:maff_relation}) to obtain an estimate of $\mbox{plim}(\widehat{MAFF}_{OR})$. We call this estimate the adjusted semiparametric method. The true MAFF is fixed as 0.5 across the simulation study; the true model is the first scenario described in Section~\ref{s:sim} with sample size $n=500$ and endemicity 0.2.

\begin{table}
\centering
\caption{Means of estimates of the MAFF in Situation 1, 2 and 3 with 1000 simulations: Neither fever killing nor measurement error (Situation 1) No fever killing, but measurement error (Situation 2) and 50\% fever killing and measurement error (Situation 3), True MAFF is 0.5.}
\begin{tabular}{c c c c c}
\hline
& S & P & L & LI \\
\hline
Situation 1 & 0.499 & 0.449 & 0.470 & 0.507\\
Situation 2 & 0.469 & 0.442 & 0.386 & 0.476 \\
Situation 3 & 0.334 & 0.286 & 0.258 & 0.370 \\
\hline
\end{tabular}

\medskip
\label{tab:compare}
\end{table}

Table~\ref{tab:compare} shows the performance of the four estimators in the three situations. In Situation 1, the adjusted semiparametric method (S) and the nonparametric method (LI) produce estimates that are approximately unbiased; however, the other two methods produce biased estimates. This biased estimation for P and L is because the two methods rely on certain model assumptions and the true model in the simulation does not satisfy these model assumptions. In Situation 2 and 3, all estimators are significantly biased from the true value 0.5. In Situation 2, the increased magnitude of biases compared to Situation 1 can be understood as biases caused by measurement error. Also, the further increase in magnitude of bias in Situation 3 compared to Situation 2 represent biases caused by 50\% fever killing. The combination of both fever killing and measurement error severely degrades the performance of the existing methods. Although the nonparametric method provides a good estimate of the MAFF in the absence of fever killing and measurement error, it performs poorly in the presence of both problems. The existing methods fail to provide unbiased estimates of the MAFF when Assumption~\ref{assumption3} is violated. 

Both the fever killing effect and measurement error are obstacles to obtain accurate measures of the parasitological challenge $\Dno$ faced by a child. The failure to measure $\Dno$ makes estimation of either $P(Y=1|D^{no.nmi})$ or $f(D^{no.nmi}|Y=1)$ biased, thus resulting in a biased estimate of the MAFF as can be seen in Table~\ref{tab:compare}. In Section~\ref{s:sim}, more simulation results are displayed in various simulation settings. In the following section, we propose our new estimation method to account for both fever killing and measurement error by considering how to recover $\Dno$ from $\Dobs$.

\section{NEW ESTIMATION METHOD}
\label{s:model}

\subsection{Notation, Data structure and the Bayes deconvolution problems}
\label{ss:bayes}

We assume that the size of the fever killing effect and the measurement error mechanism are known to researchers; if these are unknown, they can be varied in a sensitivity analysis. To model the fever killing effect, we use a parameter $\beta$ for the size of the effect; if $\beta=0.1$, then the non-malaria caused fever kills 90\% of parasites. Fever killing occurs only when $Y_i^{nmi}=1, Y_i^{mi}=0$. Therefore, $\Dcur$ is equal to $\beta \Dno$ if $Y_i^{nmi}=1, Y_i^{mi}=0$, and $\Dcur$ is the same as $\Dno$ otherwise. 

For measurement error, let $f$ be the measurement error mechanism. $\Dobs$ has the density function $f$ with the parameter $\Dcur$, i.e. $\Dobs| \Dcur \sim f(x; \Dcur)$. For example, one way to measure malaria parasites is to count the number of parasites in a fixed volume of sampled blood \citep{earle1932enumeration}; this measurement method was advocated by \citet{mckenzie2005white}. If this measurement error method is used and the only source of measurement error is the sampling of blood, then $D^{obs} | D^{cur} \sim Poisson (D^{cur})$. 

 The assumptions of a known $100(1-\beta)$\% fixed size of fever killing and known measurement error mechanism $f$ allow us to estimate the MAFF through a Bayes deconvolution approach under Assumptions~\ref{assumption2} and \ref{assumption1}. We do not assume the implausible assumption~\ref{assumption3} of no errors. The observed data $(Y_i^{obs}, D_i^{obs})$ can be split into a febrile sample ($\Yobs=1$) and an afebrile sample ($\Yobs=0$). We define
\begin{equation}
\begin{array}{ccc}
\Dno=x | \Ymi=0 &\sim & g_1(x) \\
\Dno=x | \Ymi=1 &\sim & g_2(x).
\end{array}
\label{eqn:model_dno1}
\end{equation}
where $g_1$ and $g_2$ are density functions. Given $\Ymi=1$, a child $i$ must have a positive number of parasites in her blood. Therefore, $g_2(0)$ must be zero. This restriction is considered in our estimation method in Section~\ref{ss:estimation}. Since the variables $\Ynmi$ and $\Ymi$ are independent from Assumption~\ref{assumption1} (iii), we have $\Dno=x | \Ymi=0, \Ynmi=0 \sim g_1(x)$ and  $\Dno=x | \Ymi=0, \Ynmi=1 \sim  g_1(x)$. Similarly, $\Dno=x | \Ymi=1, \Ynmi=0 \sim  g_2(x)$ and $\Dno=x | \Ymi=1, \Ynmi=1 \sim  g_2(x)$. Therefore, the $\Dno$ variable for the febrile and afebrile populations are distributed as
\begin{equation}
\begin{array}{ccc}
\Dno=x | \Yobs=0 &\sim & g_1(x) \\
\Dno=x | \Yobs=1 &\sim & (1-\lambda^*)g_1(x) + \lambda^* g_2(x).
\end{array}
\label{eqn:model_dno2}
\end{equation}
where $\lambda^*$ is the mixing proportion $\{P(\Ymi=1, \Ynmi=0)+P(\Ymi=1, \Ynmi=1)\}/P(\Yobs=1)=P(\Ymi=1)/P(\Yobs=1)$. Next, by considering the $100(1-\beta)\%$ fever killing effect, the $\Dcur$ variable is given by
\begin{equation}
\begin{array}{ccc}
\Dcur=x | \Yobs=0 &\sim & g_1(x) \\
\Dcur=x | \Yobs=1 &\sim & (1-\lambda^*)g_1^*(x) + \lambda^* g_2(x).
\end{array}
\label{eqn:model_dcur}
\end{equation}
where $g_1^* (x) = g_1 (x/\beta )$. Finally, using the measurement error mechanism $f$, the $\Dobs$ variable is distributed as
\begin{equation}
\begin{array}{ccc}
\Dobs=x | \Yobs=0 &\sim & (f \circ g_1)(x) \\
\Dobs=x | \Yobs=1 &\sim & (1-\lambda^*)(f \circ g_1^*)(x) + \lambda^* (f \circ g_2)(x).
\end{array}
\label{eqn:model3}
\end{equation}
The composite functions $(f \circ g_1^*)(x)$ and $(f \circ g_2)(x)$ can be observed and estimated from data. $\Dno$ is a random sample from unknown density $g_1$ (or $g_2$). $\Dno$ independently produces an observed random variable $\Dobs$ according to a known density $f$. When the density $f$ is known, estimating the prior densities $g_1^*$ and $g_2$ from the composite functions $(f \circ g_1^*)$ and $(f \circ g_2)$ is often called the Bayes deconvolution problem \citep{efron2016empirical}.  We note that the usual additive measurement error problem is a special case of the Bayes deconvolution problem. 

 To tackle the Bayes deconvolution problem, we apply the exponential family ``g-modeling'' approach in \citet{efron2016empirical} to the malaria data. The g-modeling approach assumes that the distributions $g_1(x)$ and $g_2(x)$ belong to exponential family distributions. Efron shows that parametric exponential family modeling can give useful estimates in moderate-sized samples while traditional asymptotic calculations are discouraging, indicating very slow nonparametric rates of convergence. Although this approach assumes a parametric model, it can be made more flexible by increasing the number of parameters as the sample size increases. With enough parameters, the risk of model misspecification is reduced.

Another approach to the Bayes deconvolution problem is to model the densities $(f \circ g_1)(x)$ and $(f \circ g_2)(x)$ directly (called ``f-modeling'' in \citet{efron2014two} ). The f-modeling approach is not considered in this article because the g-modeling approach is more straightforward to capture the fever killing effect. Knowing $g_1(x)$  implies knowing $g_1(\beta x)$, but this is not true in the f-modeling approach. Furthermore, g-modeling is more efficient than f-modeling \citep{efron2016empirical}.

\subsection{Estimation}
\label{ss:estimation}

First, we discretize the $D^{cur}$ space for the afebrile sample and the febrile sample separately. To be specific, considering the $100(1-\beta)$\% fever killing effect, we define the discrete sets of $D_a^{cur}$ and $D_f^{cur}$ for the afebrile population and febrile population respectively;
\[
D_a^{cur} = \{d_1, ..., d_k \}, D_f^{cur} = \{\beta d_1, ..., \beta d_k \}.
\]
Since the parasite level is always greater than or equal to zero, the element $d_1$ of the discrete set is zero. We note that we can use continuous formulation instead of using discretization. However, this discretization is essentially necessary because the numerical implementation of the theory is required to obtain the maximum likelihood estimates.

The g-modeling approach assumes that the distributions $g_1, g_1^*$ and $g_2$ are the exponential family distributions. Also, we assume that both distributions $g_1$ and $g_1^*$ share the common parameters because of $g_1(x) = g_1^*(\beta x)$. Assume that the densities of $g_1, g_1^*$ and $g_2$ are
\begin{equation}
\begin{array}{ccll}
g_1(d_j ; \alpha_1) &=& \exp\{Q_{1j} \alpha_1 - \phi_1(\alpha_1)\} &, \> j=1, ..., k \\
g_1^*(\beta d_j ; \alpha_1) &=& g_1(d_j ; \alpha_1) &, \> j=1, ..., k \\
g_2(\beta d_j ; \alpha_2) &=& \exp\{Q_{2j} \alpha_2 - \phi_2(\alpha_2)\} &, \> j=1, ..., k \\
\end{array}
\label{g_model}
\end{equation}
where (1) $\alpha_1$ is a $m_1$-dimensional parameter vector and $\alpha_2$ is a $m_2$-dimensional parameter vector, (2) $Q_{1j}$ and $Q_{2j}$ are $j$-th rows of the known matrices $Q_{1}$ and $Q_{2}$ respectively and (3) $\phi_1(\alpha_1)$ and $\phi_2(\alpha_2)$ are chosen to satisfy $\sum_{j=1}^{k} g_1(d_j; \alpha_1) =  \sum_{j=1}^{k} g_2(\beta d_j; \alpha_2) = 1$. $Q_1$ is a $k \times m_1$ matrix and $Q_2$ is a $k \times m_2$ matrix. The matrices $Q_1$ and $Q_2$ are determined by the sample spaces $D_a^{cur}$ and $D_f^{cur}$ respectively, and they can be constructed by using the natural spline R function `ns'; the matrix $Q_1$ is obtained from `ns($D_a^{cur}$, df=$m_1$)', and the matrix $Q_2$ is obtained from `ns($D_f^{cur}$, df=$m_2$)'. Then, the matrices are standardized so that each column has mean zero and sum of squares one. For further computational details, see \citet{efron2016empirical}.

 Given the density functions, the probability of having an observation $x_i$ in the afebrile sample is 
\[
P(D_i^{obs} = x_i | \Yobs=0) = \sum_{j=1}^{k} P(D_i^{obs} = x_i | D_i^{cur} = d_j) g_1(d_j; \alpha_1) = \sum_{j=1}^{k} f(x_i; d_j) g_1(d_j; \alpha_1) 
\]
and in a similar fashion, the probability of having an observation $x_i$ in the febrile sample is 
\begin{equation}
\begin{array}{ccl}
P(D_i^{obs} = x_i | \Yobs=1) &=& \sum_{j=1}^{k} P(D_i^{obs} =  x_i | D_i^{cur} = \beta d_j) \cdot \{(1-\lambda^*)g_1^*(\beta d_j; \alpha_1) + \lambda^* g_2(\beta d_j; \alpha_2)\}\\
&=& \sum_{j=1}^{k} (1-\lambda^*)f(x_i; \beta d_j)g_1^*(\beta d_j; \alpha_1) + \lambda^* f(x_i; \beta d_j)g_2(\beta d_j; \alpha_2).
\end{array}
\label{prob_f}
\end{equation}
Then, the likelihood $\mathcal{L}(p, \lambda^*, \alpha_1, \alpha_2)$ is written as
\begin{equation}
\begin{array}{ccl}
\mathcal{L}(p, \lambda^*, \alpha_1, \alpha_2) &=& \prod_{i=1}^{n} P(\Dobs=x_i, \Yobs=y_i) \\
&=& \prod_{i=1}^{n} \bigg[ \{ P(D_i^{obs} = x_i|\Yobs=1) P(\Yobs=1)\}^{I(\Yobs=1)}\\
&& \quad\quad \times \{P(D_i^{obs} = x_i | \Yobs=0)P(\Yobs=0)\}^{I(\Yobs=0)} \bigg]. 
\end{array}
\label{likeli}
\end{equation}
and the log-likelihood $\ell(p, \lambda^*, \alpha_1, \alpha_2)$ is written as
\begin{equation}
\begin{array}{ccl}
\ell(p, \lambda^*, \alpha_1, \alpha_2) &=& \sum_{i=1}^{n} I(\Yobs=0) \cdot \{ \log(1-p) + \log P(D_i^{obs} = x_i| \Yobs=0; \alpha_1) \} \\
&& \quad\>\> + I(\Yobs=1) \cdot \{ \log{p} + \log P(D_i^{obs} = x_i | \Yobs=1; \lambda^*,\alpha_1, \alpha_2)\} \\
&=& \sum_{i=1}^{n} I(\Yobs=0) \cdot \log(1-p) + I(\Yobs=1) \cdot \log{p} \\
&& \quad\>\> + I(\Yobs=0) \cdot \log\{ \sum_{j=1}^{k} f(x_i; d_j) g_1(d_j; \alpha_1)\}\\
&& \quad\>\> + I(\Yobs=1) \cdot \log\{ \sum_{j=1}^{k} (1-\lambda^*)f(x_i; \beta d_j)g_1^*(\beta d_j; \alpha_1) + \lambda^* f(x_i; \beta d_j)g_2(\beta d_j; \alpha_2) \}. 
\end{array} 
\label{log_likeli}
\end{equation}
The estimates $\hat{\lambda}^*$ and $\hat{p}$ can be obtained by maximizing the log-likelihood $\ell(p, \lambda^*, \alpha_1, \alpha_2)$. Then, the estimate $\hat{\lambda}$ of the MAFF is obtained from $\hat{\lambda}^*$ and $\hat{p}$, $\hat{\lambda} = (\hat{\lambda}^*-\hat{p}\hat{\lambda}^*)/(1-\hat{p}\hat{\lambda}^*)$ by using adjustment~(\ref{eqn:maff_relation}). 

Also, \citet{efron2016empirical} suggests to use a penalized likelihood as he finds that the accuracy of a deconvolution estimate obtained from the g-modeling approach can be greatly improved by regularization of the maximum likelihood algorithm. Instead of maximizing $ \ell(p, \lambda^*, \alpha_1, \alpha_2)$, we maximize a penalized log-likelihood 
\[
m(p, \lambda^*, \alpha_1, \alpha_2) = \ell(p, \lambda^*, \alpha_1, \alpha_2) - s(\alpha_1, \alpha_2)
\]
where $s(\alpha_1, \alpha_2)$ is a penalty function. In this article, the function $s(\alpha_1, \alpha_2)$ is 
\[
s(\alpha_1, \alpha_2) = c_0(||\alpha_1||^2 + ||\alpha_2||^2)^{1/2}
\]
where $c_0$ is a regularizing constant. The choice of $c_0$ is related to bias-variance tradeoff. \citet{efron2016empirical} suggests $c_0=1$ as a modest choice that restricts the trace of the added variance due to the penalization. 

For the computation, an equispaced discretization is used. The `optim' R package is used to find the maximum likelihood estimates by restricting $p$ and $\lambda^*$ inside the interval (0,1). This restriction can be done by using `L-BFGS-B' method option in the optim function. The R code file for this estimation approach is provided in the Supplementary Materials. 

\subsection{Sensitivity Analysis for Assumption~\ref{assumption1}}
\label{ss:violation}

Our proposed estimation method relies on Assumption~\ref{assumption1}, particularly, Assumption~\ref{assumption1} (iii) and (iv) that assume $\Ymi$ and $\Ynmi$ have independent causal pathways that form a parallel circuit to trigger a fever, in order to deal with the mixture problem. However, these assumptions could potentially be violated. For example, if a child has a non-malarial infection, but it is not strong enough to trigger a non-malarial fever ($Y_i^{nmi}=0$), then the parasite density just below the threshold ($Y_i^{mi}=0$) can trigger a fever ($Y_i^{obs}=1$) because a combined effect of some malaria parasites and some non-malarial infections that make the child's immune system weaker might be enough to trigger a fever. In this scenario that violates Assumption~\ref{assumption1} (iv), if all the malaria parasites in the child's blood are removed, then the non-malarial infection will be still below the threshold $T_i^{nmi}$, and the observed fever $Y_i^{obs}$ will be removed. Therefore, the child has a fever attributable to malaria, but $Y_i^{mi}$ does not correctly describe this malaria attributable fever.

\begin{figure}
\centering
\includegraphics[scale=0.66]{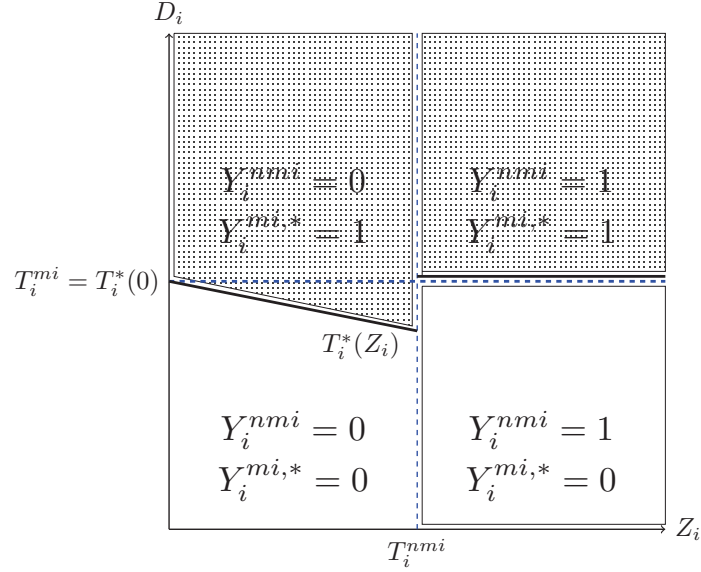}
\caption{Comparison of separation by $Y^{mi}$ and $Y^{nmi}$ and separation by $Y^{mi,*}$ and $Y^{nmi}$.}
\label{fig_separation}
\end{figure}

To propoerly describe this scenario, we instead consider a modified pyrogenic threshold variable $T_i^*(Z_i)$ which is a decreasing function of the non-malarial infection level $Z_i$ and $T_i^*(0)=T_i^{mi}$ and consider a new variable $\Ymistar$, which is $I(D_i \geq T_i^*(Z_i))$ if $Z_i < \Tnmi$ and is $I(D_i \geq  T_i^*(0))$ if $Z_i \geq \Tnmi$. Figure~\ref{fig_separation} displays the difference between $\Ymi$ and $\Ymistar$. $\Ymi=1$ indicates the area above the horizontal dashed line and $\Ymistar=1$ indicates the dotted area above the solid line. $\Ymi$ is equal to $\Ymistar$ if $Z_i \geq \Tnmi$ by definition. We note that the definition of the MAFF does not depend on how to separate $Y^{nmi}=1, Y^{mi}=1$ and $Y^{nmi}=1, Y^{mi}=0$ because the MAFF only depends on the proportion of $Y^{nmi}=0$ given $Y^{obs}=1$. Therefore, $Y^{mi}$ has no role in defining the MAFF. Like $\Ymi$, $\Ymistar$ is a nuisance variable and is used to separate the area of $\Ynmi=1$.

We can use the same estimation method described in Section~\ref{ss:estimation} if $\Ynmi$ and $\Ymistar$ satisfy Assumption~\ref{assumption1}. However, both $\Ynmi$ and $\Ymistar$ depend on the amount of non-malarial infections $Z_i$, and they are dependentm which means Assumption~\ref{assumption1} (iii) would be violated with $\Ymistar$ in place of $\Ymi$. One way to mitigate this violation of the independence assumption of $Y_i^{nmi}$ and $Y_i^{mi,*}$ is to conduct a sensitivity analysis. The violation affects two relationships: the relationship between the parasite densities $f(\Dno=x | \Ymistar=0, \Ynmi=0)$ and $f(\Dno=x | \Ymistar=0, \Ynmi=1)$ and the relationship between the probabilities $P(\Ynmi=0 | \Ymistar=1)$ and $P(\Ynmi=0 | \Ymistar=0)$. Under the independence assumption, the two densities are equal and the two probabilities are equal. However, if the independence assumption does not hold, then the relationships are not identified. To identify the MAFF, we need sensitivity parameters to describe and restrict the relationships.

The relationship between $f(\Dno=x | \Ymistar=0, \Ynmi=0)$ and $f(\Dno=x | \Ymistar=0, \Ynmi=1)$ can be described by using an exponential tilt model,  
\[
f(\Dno=x | \Ymistar=0, \Ynmi=1) = \exp(\delta_0 + \delta_1 x)f(\Dno=x | \Ymistar=0, \Ynmi=0).
\]
where $\delta_0$ and $\delta_1$ are known. If $\delta_1$ is known, $\delta_0$ is identified because of $\int_0^{\infty} f(\Dno=x | \Ymistar=0, \Ynmi=1) =1 \> dx$, and is a function of $\delta_1$. We use $\delta_1$ as a sensitivity parameter. The g-modeling method assumes the parasite density $f(\Dno=d_j | \Ymistar=0, \Ynmi=0) = \exp\{Q_{1j} \alpha_1 - \phi_1(\alpha_1) \}$ where $\phi_1(\alpha_1) = \log \big(\sum_{j=1}^{k} \exp\{Q_{1j} \alpha_1 \} \big)$. Therefore, the parasite density $f(\Dno=d_j | \Ymistar=0, \Ynmi=1)$ is represented as
\[
\exp(\delta_0 + \delta_1 d_j) \exp\{Q_{1j} \alpha_1 - \phi_1(\alpha_1) \} = \exp\{(\delta_1 d_j + Q_{1j} \alpha_1)+ (\delta_0 - \phi_1(\alpha_1)) \},
\]
and $\delta_0$ is chosen as $\delta_0 = \log \big(\sum_{j=1}^{k} \exp\{Q_{1j} \alpha_1 \} \big) - \log \big(\sum_{j=1}^{k} \exp\{Q_{1j} \alpha_1 +\delta_1 d_j\} \big)$. From the definition of $\Ymistar$, it is evident that the proportion of low parasites when $\Ymistar=0, \Ynmi=0$ is always higher than that when $\Ymistar=0, \Ynmi=1$, which can be seen visually in Figure~\ref{fig_separation}. This restricts the sensitivity parameter $\delta_1$ to satisfying the inequality $\delta_1 \geq 0$. For the relationship between $P(\Ynmi=0 | \Ymistar=1)$ and $P(\Ynmi=0 | \Ymistar=0)$, we use a sensitivity parameter $\tau$ as the ratio of the probabilities $\tau= P(\Ynmi=0 | \Ymistar=1)/P(\Ynmi=0 | \Ymistar=0)$. From the definition of $\Ymistar$, $\tau$ should be above 1, i.e. $\tau \geq 1$. We conduct a sensitivity analysis by using the sensitivity parameter $\delta_1$ and $\tau$ in Section~\ref{s:example} to investigate the impact of the violation of Assumption~\ref{assumption1} (iii) on the estimate of the MAFF. 

The independence assumption can also be violated when $\Ynmi$ and $\Dno$ are dependent because of confounding variables. If these confounders are observed covariates such that $Y^{nmi}$ and $D^{no.nmo}$ are conditionally independent given these observed covariates, then we can estimate the MAFF by extending our proposed method to incorporate the observed covariate. Let the observed covariates be denoted by the $r$-dimensional vector $O_i$. In the presence of covariates, rather than allowing any function for $f$, we assume a one-parameter exponential family of conditional densities for each $O_i$; $f(\Dno | \Dcur, O_i) = \exp{\{ \Dno \cdot \eta_i  - \psi(\eta_i) \}}$
where $\eta_i=\Dcur + O_i^T \gamma$ is a linear combination of $\Dcur$ and $O_i$, and $\gamma$ is an unknown $r$-dimensional parameter vector \citep{efron2016empirical}. The functional form is known, but the parameter vector $\gamma$ is not. Therefore, $\gamma$ has to be estimated in the estimation procedure. In addition to a unknown $m$-dimensional parameter vector $\alpha$ for $g(\Dcur; \alpha)$, $(\alpha, \gamma)$ can be estimated by maximizing the likelihood function as we discussed in Section~\ref{ss:estimation}.

\section{SIMULATION STUDY}
\label{s:sim}

In this section, we evaluate the performance of our proposed method including the regular likelihood approach and the penalized likelihood approach in a simulation study. It was shown in Section~\ref{ss:simul_methods} that the existing methods produce biased estimates of the MAFF when there is fever killing and measurement error. In addition to evaluating the performance of our proposed method, we compare it to the existing methods by considering various simulation settings. The distribution of $g_1$ is a mixture of a point mass at zero and a distribution for positive parasite levels and the distribution of $g_2$ satisfies $g_2(0)=0$. We consider two scenarios; (i) $g_1$ and $g_2$ are exponential family distributions and (ii) $g_1$ and $g_2$ are not exponential family distributions. For the first scenario, we assume
\[
g_1(x) = q \cdot I(x=0) +(1-q) \cdot \mbox{TN}_{(0,\infty)}(\mu_1, \sigma_1)
\]
where $q$ is the proportion of zero parasite level and $\mbox{TN}_{(0,\infty)}(\mu, \sigma)$ is a truncated normal distribution with mean $\mu$ and standard deviation $\sigma$ in the interval $(0, \infty)$. The distribution of $g_1^*(x)$ is implicitly defined as 
\[
g_1^*(x) = q \cdot I(x=0) +(1-q) \cdot \mbox{TN}_{(0,\infty)}(\beta \mu_1, \beta \sigma_1), 
\]
and the distribution of $g_2$ can only take positive parasite levels,
\[
g_2(x) =  \mbox{TN}_{(0,\infty)}(\mu_2, \sigma_2).
\]
The second scenario considers uniform distributions, which are not in exponential family distributions. To be specific, $g_1(x) = q_1 \cdot I(x=0) +(1-q_1)q_2 \cdot \mbox{TN}_{(0,\infty)}(\mu_1, \sigma_1) + (1-q_1)(1-q_2) \cdot U(0, 2\mu_1)$ and $g_2(x) =  q_2 \cdot \mbox{TN}_{(0,\infty)}(\mu_2, \sigma_2) + (1-q_2) \cdot U(0, 2\mu_2)$ where $U(a, b)$ is the uniform distribution in the interval $(a, b)$. Throughout the simulation study, the probability $q_1$ is fixed as 1/8. The number of the parameters is chosen as $m_1=4$ for $g_1$ and $m_2=3$ for $g_2$. Also, we assume that we know the Poisson measurement error mechanism $D^{obs} \sim Pois(D^{cur})$ i.e. the mechanism $f(x)$ is the standard Poisson distribution. 

\begin{table}[t]
\centering
\caption{Exponential family distribution case. Means (standard deviations) of the estimators in simulation settings are displayed; P represents the power model regression method, S represents the adjusted semiparametric method, and LI represents the nonparametric method. True MAFF is 0.5}
\resizebox{\textwidth}{!}{%
\begin{tabular}{ccccc ccc}
\toprule
& & & \multicolumn{5}{c}{MAFF} \\
\cmidrule{4-8} 
$n$ & $q$ & $\beta$ & Regular & Penalized & P& S & LI\\
\midrule
500 & 0.2 & 1 & 0.501 (0.120) & 0.488 (0.118) & 0.443 (0.089) & 0.471 (0.083)& 0.475 (0.078) \\
& & 0.8 & 0.499 (0.117) & 0.487 (0.112) & 0.406 (0.102) & 0.437 (0.089)& 0.441 (0.081) \\
& & 0.2 & 0.512 (0.056) & 0.491 (0.045) & 0.119 (0.037) & 0.138 (0.064)& 0.143 (0.059) \\
& 0.8 & 1 & 0.499 (0.028) & 0.497 (0.028) & 0.485 (0.027) & 0.487 (0.026)& 0.483 (0.025) \\
& & 0.8 & 0.498 (0.028) & 0.495 (0.028) & 0.483 (0.027) & 0.485 (0.027)& 0.480 (0.025) \\
& & 0.2 & 0.499 (0.024) & 0.495 (0.022) & 0.453 (0.025) & 0.457 (0.024)& 0.444 (0.020) \\
1000 & 0.2 & 1 & 0.499 (0.081) & 0.490 (0.082) & 0.440 (0.063) & 0.468 (0.058)& 0.472 (0.055)\\
& & 0.8 & 0.500 (0.076) & 0.490 (0.072) & 0.406 (0.072) & 0.437 (0.062)& 0.438 (0.057)\\
& & 0.2 & 0.507 (0.039) & 0.495 (0.031) & 0.117 (0.022) & 0.138 (0.039)& 0.135 (0.042) \\
& 0.8 & 1 & 0.498 (0.020) & 0.496 (0.020) & 0.484 (0.019) & 0.486 (0.019)& 0.482 (0.018)\\
& & 0.8 & 0.498 (0.020) & 0.497 (0.020) & 0.483 (0.019) & 0.485 (0.018)& 0.480 (0.017)\\
& & 0.2 & 0.499 (0.017) & 0.497 (0.016) & 0.455 (0.018) & 0.458 (0.017)& 0.446 (0.015) \\
\bottomrule
\end{tabular}
}
\medskip
\label{tab:sim}
\end{table}

In addition, we consider the following three factors that may affect the performance of our method. 

\begin{enumerate}
\item Size of fever killing effect. Three different sizes of the fever killing effect are considered. The settings are large fever killing effect (fever kills approximately 80\% of parasites which means $\beta = 0.2$), small fever killing effect (fever kills approximately 20\% of parasites, $\beta=0.8$) and no fever killing effect ($\beta=1$). The no fever effect case will be used as a standard for comparison between the other two fever killing effects settings. 

\item Endemicity. Endemic regions differ greatly by how many people have the malaria parasites in their blood. Endemicity could affect the variance of the estimate of the MAFF. Two levels of endemicity are considered: mesoendemic $q=0.8$ (low to moderate) and holoendemic $q=0.2$ (high). 

\item Sample size $n$. Two sample sizes, $n=500$ and $n=1000$, are considered. 

\end{enumerate}
There are $3 \times 2 \times 2 =12$ settings for each scenario to investigate the effect of the settings on the performance of our proposed method.

\begin{table}[t]
\centering
\caption{Non-exponential family distribution case. Means (standard deviations) of the estimators in simulation settings are displayed; P represents the power model regression method, S represents the adjusted semiparametric method, and LI represents the nonparametric method. True MAFF is 0.5}
\resizebox{\textwidth}{!}{%
\begin{tabular}{ccccc ccc}
\toprule
& & & \multicolumn{5}{c}{MAFF} \\
\cmidrule{4-8} 
$n$ & $q$ & $\beta$ & Regular & Penalized & P & S & LI\\
\midrule
500 & 0.2 & 1 & 0.492 (0.116) & 0.499 (0.126) & 0.423 (0.094)& 0.458 (0.087)& 0.466 (0.081)\\
& & 0.8 & 0.488 (0.112) & 0.503 (0.121) & 0.379 (0.103)& 0.420 (0.091)& 0.430 (0.080)\\
& & 0.2 & 0.516 (0.062) & 0.517 (0.071) & 0.123 (0.033)& 0.133 (0.060)& 0.140 (0.055)\\
& 0.8 & 1 & 0.498 (0.029) & 0.501 (0.029) & 0.480 (0.027)& 0.482 (0.027)& 0.478 (0.026)\\
& & 0.8 & 0.497 (0.030) & 0.499 (0.031) & 0.479 (0.027)& 0.481 (0.026)& 0.475 (0.025)\\
& & 0.2 & 0.501 (0.025) & 0.505 (0.027) & 0.451 (0.025)& 0.454 (0.024)& 0.441 (0.021)\\
1000 & 0.2 & 1 & 0.502 (0.091) & 0.499 (0.095) & 0.420 (0.065)& 0.454 (0.060)& 0.462 (0.055)\\
& & 0.8 & 0.498 (0.090) & 0.498 (0.093) & 0.379 (0.074)& 0.421 (0.063)& 0.430 (0.056)\\
& & 0.2 & 0.503 (0.051) & 0.493 (0.040) & 0.125 (0.022)& 0.138 (0.038)& 0.137 (0.040)\\
& 0.8 & 1 & 0.497 (0.020) & 0.497 (0.020) & 0.481 (0.019)& 0.483 (0.018)& 0.478 (0.018)\\
& & 0.8 & 0.498 (0.022) & 0.498 (0.022) & 0.480 (0.019)& 0.482 (0.019)& 0.476 (0.018)\\
& & 0.2 & 0.501 (0.019) & 0.498 (0.020) & 0.450 (0.017)& 0.453 (0.017)& 0.440 (0.015)\\
\bottomrule
\end{tabular}}
\medskip
\label{tab:sim2}
\end{table}

We use both the regular likelihood approach and the penalized likelihood approach to estimate the MAFF and compare them to the existing methods that do not account for fever killing and measurement error. Table~\ref{tab:sim} shows the means and standard deviations of the maximum likelihood estimates in the various settings when the true models are in the exponential family. The means and the standard deviations of the estimates are obtained from 1000 repetitions. Three aspects are found in this table. First, both the regular likelihood and the penalized likelihood approaches provide approximately unbiased estimates of the true MAFF while other existing methods (P, S, LI) are biased. There is a trend that the regular likelihood estimates have lower bias and slightly larger standard deviations than the penalized likelihood estimates. This is because the true models are exponential family distributions. Second, the larger proportion of zero parasite level contributes to more efficient estimates. That is, the estimate of the MAFF in mesoendemic regions is more efficient than the estimate in holoendemic regions. Finally, as sample size $n$ increases, both the approaches produce estimates which are closer to the true MAFF value 0.5 (less bias) and have smaller standard deviations. 

Table~\ref{tab:sim2} displays the means and standard deviations of the estimates when the true models are not exponential family distributions. As can be seen in Table~\ref{tab:sim2}, a larger $n$ and a larger $q$ contribute to a smaller standard deviation. A smaller $\beta$ leads to a smaller standard deviation, but it also leads to a bias. A different aspect in Table~\ref{tab:sim2} compared to Table~\ref{tab:sim} is that the penalized likelihood approach generally performs better with smaller standard deviations than the regular likelihood approach in the non-exponential family distribution case of Table~\ref{tab:sim2}.


\section{APPLICATION: DATA FROM KILOMBERO, TANZANIA}
\label{s:example}

\begin{table}
\centering
\caption{Summary of the data from Kilombero, Tanzania}
\begin{tabular}{ccc}
\hline
Parasite level & Afebrile  & Febrile  \\
\hline
$=0$ & 160 & 16\\
$>0$ & 1698 & 121 \\
\hline
Total & 1858 & 137 \\
\hline
\end{tabular}
\medskip
\label{tab:data}
\end{table}

We consider data from a study of children in the Kilombero District (Morogoro Region) of Tanzania \citep{smith1994attributable}. The study collected parasite density levels and the presence of fever among 426 children under six years of age in two villages from June 1989 until May 1991 in the Kilombero District (Morogoro Region) of Tanzania. This area is highly endemic for \textit{Plasmodium falciparum} malaria. A total of 1996 blood films from the 426 children were examined. \citet{smith1994attributable} found that the correlation between consecutive observations on the same child is not significant and the impact of the correlation on the MAFF is negligible. We will follow \citet{smith1994attributable} in assuming that the 1996 collected observations are independent and for brevity will describe the data as involving 1996 children. In this dataset, there are $n_1=137$ children who had a fever and $n_0=1859$ children who did not have a fever. The former is a group of febrile children whose fever was caused by either malaria infection or non-malaria infection. The latter is a group of afebrile controls that is used to provide information on the parasite density of the non-malaria infection population. Table~\ref{tab:data} summarizes the data. The proportions of zero parasite level is $0.086=160/1858$ and $0.117=16/137$ in the afebrile and febrile groups respectively. Under the no errors assumption (Assumption~\ref{assumption3}) and the monotonicity assumption in Assumption~\ref{assumption1}, the proportion of zero parasite density level in the afebrile group must be greater than that in the febrile group so the proportions in Table~\ref{tab:data} suggest some violations of the assumptions. 

\begin{figure}
\centering
\includegraphics[scale=0.6]{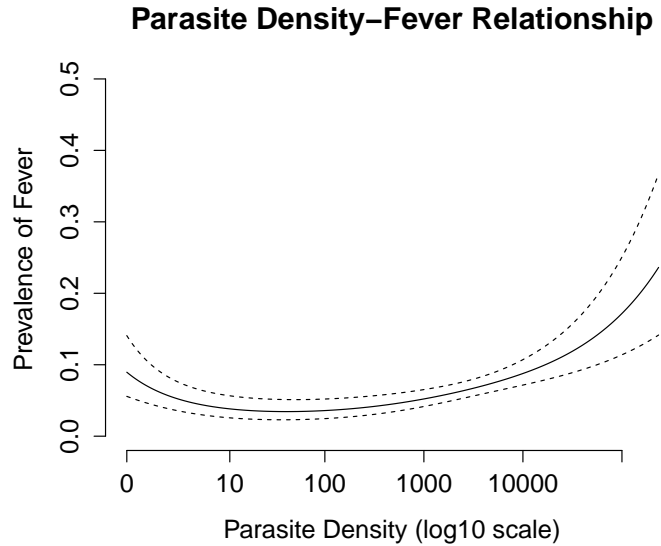}
\caption{The relationship between parasite density and probability of fever. The solid curve represents the point estimate across parasite density obtained by using penalized splines, and the dashed curves are 95\% pointwise confidence intervals. }
\label{fig_relationship}
\end{figure}

Also, in the absence of measurement error and fever killing, under Assumption~\ref{assumption1}, the probability of a fever should be monotonically increasing in the parasite density. Figure~\ref{fig_relationship} shows the relationship between probability of fever and parasite density and also suggests a violation of the assumptions. \citet{smith1994attributable} point out that this phenomenon that the probability of a fever is not monotonically increasing in the parasite density has been observed in many other datasets, and consider it a consequence of non-malarial fevers suppressing low density parasitaemia, i.e., fever killing.

The existing four estimation methods discussed earlier (L, P, IL, S discussed in Section~\ref{s:fever}) can be used to estimate the MAFF from the malaria data in Tanzania, but the methods do not account for the fever killing effect and measurement error problems that we have described in Section~\ref{s:fever}. Under the assumption of the absence of the fever killing effect and measurement error, the estimates of the MAFF from the existing methods are 0.176 (L), 0.202 (P), 0.177 (LI) and 0.177 (S) shown in the bottom of Table~\ref{tab:maff}. The displayed standard deviations are computed from 2000 bootstrapped samples. 

To apply our proposed method, a measurement error mechanism needs to be specified. In the Kilombero study, the number of parasites is counted in a predetermined number of white blood cells (WBCs), usually 200, and then the parasite density per $\mu$l is estimated as 40 times the count, under the assumption that there are 8000 WBCs per $\mu$l.  The simplest choice is the Poisson measurement error mechanism. The Poisson measurement error mechanism will hold if the only source of measurement error is the sampling of parasites from 1/40 $\mu$l of blood and parasites are uniformly distributed throughout the blood. Without any other sources of measurement error but this sampling error, the measurement error model would be (M1) $f_1(D^{obs}|D^{cur}) \sim Poisson(D^{obs}/40 ; D^{cur}/40)$. As we discussed in Section~\ref{s:fever}, microscopy error is another source of measurement error as well as sampling error. A more complicated model than the Poisson model to account for the microscopy error is a negative binomial measurement error model. The negative binomial (NB) model is (M2) $f_2(D^{obs}|D^{cur})  \sim  $ NB$(D^{obs}/40; r, r/({r+ D^{cur}/40}))$ where $r$ is the dispersion parameter. From \citet{o2007malaria}, we estimate the dispersion parameter is $r=6$. The estimation process is shown in Appendix~\ref{appx:computation}. Furthermore, another source of measurement error is WBCs count variability; the number of WBCs per $\mu\mathrm{l}$ is not fixed as 8000, but varies from person to person \citep{mckenzie2005white}. Based on \citet{mckenzie2005white}, we consider a discrete distribution $h(x)$ of WBCs counts per $\mu\mathrm{l}$: Point masses of $(4, 5, 6, 7, 8, 9, 10, 11, 12)\times 10^3=(.12, .16, .20, .16, .16, .10, .04, .04, .02)$. The distribution $h(x)$ accounts for the potential effect of the variability of WBCs counts. The most complicated measurement error model (M3) in our simulation study combines the microscopy error and the WBC count variability, and is $f_3(D^{obs}|D^{cur})  \sim \sum h(x) \cdot $ NB$(D^{obs}/(x/200); r, r/({r+ D^{cur}/(x/200)}))$. We note that the distribution $f_1$ has the smallest variance and the distribution $f_3$ has the largest variance. The models (M1)-(M3) are considered and compared in our analysis. 

\begin{table}[t]
\centering
\caption{Estimates of the MAFF. The upper table: the estimates corresponding to the different sizes of fever killing; $1-\beta=0.95$ means 95\% fever killing and $1-\beta=0$ means no fever killing. The standard deviations are computed from 500 bootstrapped samples. The lower table: the estimates from the existing methods. }
\resizebox{\textwidth}{!}{%
\begin{tabular}{cccc c@{\hskip 0.5in} cccc}
\toprule
& \multicolumn{3}{c}{MAFF} & & & \multicolumn{3}{c}{MAFF} \\
\cline{2-4} \cline{7-9}
$1-\beta$ & M1 & M2 & M3 & &$1-\beta$ & M1 & M2 & M3 \\
\midrule
0.00 & 0.200 (0.057) & 0.192 (0.056) & 0.190 (0.057) & & 0.50 & 0.302 (0.056) & 0.336 (0.054) & 0.330 (0.053)\\
0.05 & 0.209 (0.056) & 0.203 (0.050) & 0.201 (0.050) & & 0.55 & 0.317 (0.061) & 0.359 (0.057) & 0.352 (0.056) \\
0.10 & 0.217 (0.053) & 0.215 (0.048) & 0.212 (0.052) & & 0.60 & 0.335 (0.066) & 0.385 (0.064) & 0.377 (0.063)\\
0.15 & 0.227 (0.046) & 0.227 (0.052) & 0.224 (0.052) & & 0.65 & 0.357 (0.064) & 0.415 (0.067) & 0.405 (0.064)\\
0.20 & 0.236 (0.050) & 0.239 (0.052) & 0.236 (0.049) & & 0.70 & 0.388 (0.076) & 0.449 (0.068) & 0.438 (0.069)\\
0.25 & 0.245 (0.049) & 0.253 (0.054) & 0.249 (0.051) & & 0.75 & 0.428 (0.072) & 0.490 (0.075) & 0.476 (0.076)\\
0.30 & 0.255 (0.049) & 0.267 (0.054) & 0.263 (0.049) & & 0.80 & 0.482 (0.086) & 0.538 (0.083) & 0.521 (0.080)\\
0.35 & 0.265 (0.051) & 0.282 (0.050) & 0.278 (0.050) & & 0.85 & 0.557 (0.082) & 0.598 (0.082) & 0.578 (0.086)\\
0.40 & 0.276 (0.048) & 0.298 (0.050) & 0.294 (0.052) & & 0.90 & 0.660 (0.073) & 0.680 (0.073) & 0.655 (0.071)\\
0.45 & 0.288 (0.054) & 0.316 (0.055) & 0.311 (0.055) & & 0.95 & 0.805 (0.052) & 0.821 (0.053) & 0.794 (0.054)\\
\bottomrule
\end{tabular}}
\resizebox{0.7\textwidth}{!}{%
\begin{tabular}{ccccc}
\toprule
 & L & P & S & LI \\
 \cmidrule{2-5}
MAFF & 0.176 (0.042) & 0.202 (0.074)  & 0.177 (0.063) & 0.182 (0.079)\\
\bottomrule
\end{tabular}}
\label{tab:maff}
\medskip
\end{table}

\begin{figure}
\centerline{
\includegraphics[width=150mm]{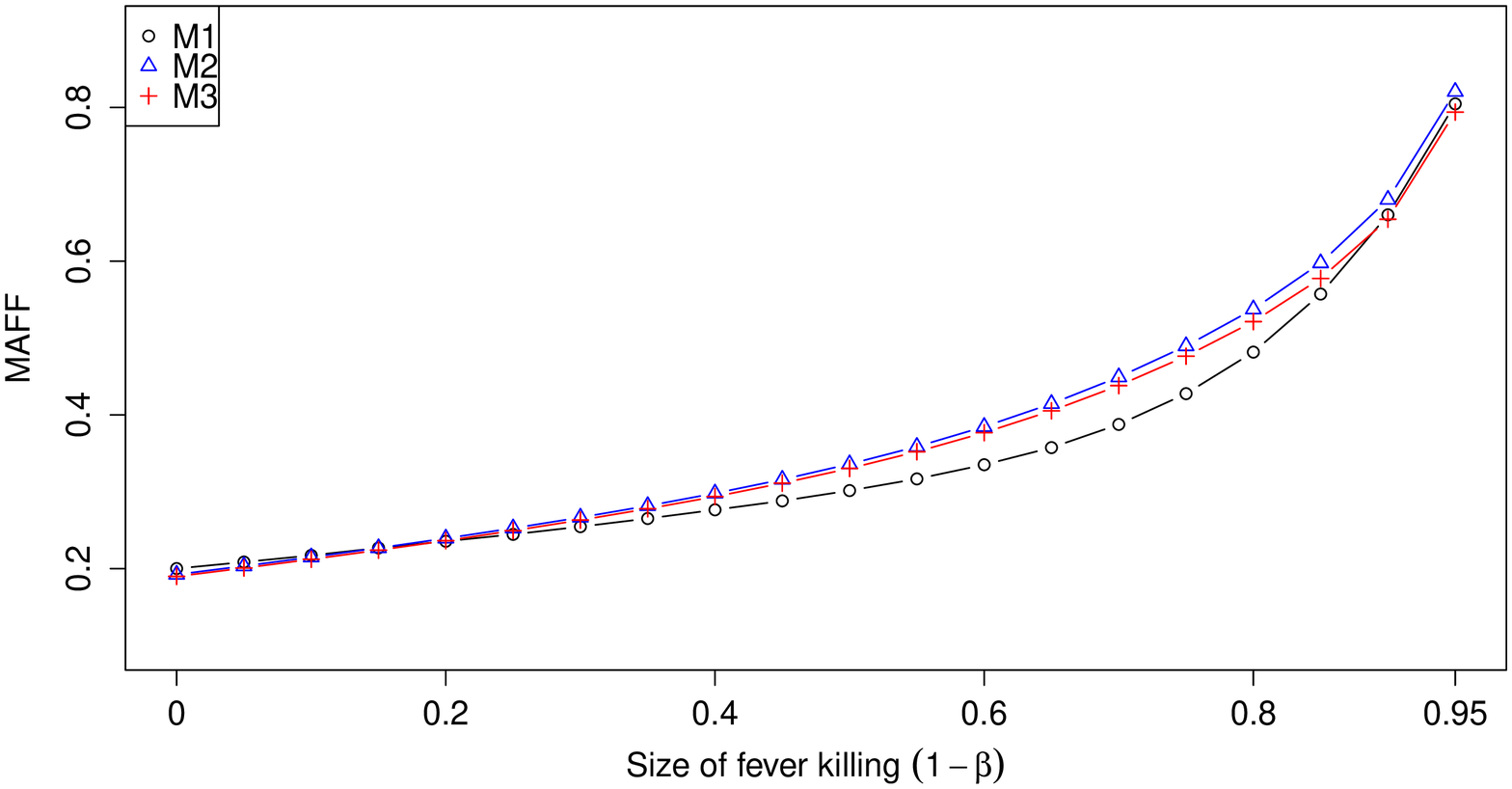}}
\caption{The plot of the estimates of the MAFF on the fever killing parameter $\beta$ for the measurement error models (M1)-(M3).}
\label{fig:maff}
\end{figure}

In addition to the measurement error mechanism, the size of fever killing $1- \beta$ needs to be specified, but it is not known precisely based on current scientific knowledge. We consider a series of various plausible fever killing sizes, and calculate the corresponding estimates of the MAFF; the series of fever killing sizes ranges from no fever killing to 95\% fever killing in the increment 0.05, i.e., the sequence of $\beta$ is $(0.05, 0.10, \ldots, 0.95, 1.00)$. We do not consider no fever killing to be plausible but include it for comparison purposes. 

Based on previous studies, we can shorten the plausible range of the size of fever killing effect by using children's temperature data. We found that the temperature data are distributed between $37.5^{\circ}$C and $40^{\circ}$C with 90\% percentile $38.7^{\circ}$C and 95\% percentile $39.1^{\circ}$C. \citet{long2001plasmodium} found that when the temperature is $38.8^{\circ}$C, the fever killing effect was 50\% so we consider 50\% an upper bound on the fever killing effect (i.e., the range of the fever killing size is $1-\beta \in (0, 0.5)$). Furthermore, the assumption that has a fixed size of fever killing across population can be eased by incorporating the temperature data into the analysis by accounting for temperature-varying fever killing size. However, incorporating the temperature data is beyond the scope of our paper.  

Table~\ref{tab:maff} shows the estimates of the MAFF from the different values of the fever killing effect parameter $\beta$ for each measurement error model from (M1) to (M3). As the fever killing effect becomes larger (i.e., $\beta$ decreases), the estimate of the MAFF increases from 0.200 to 0.805 of the model (M1), the estimate increases from 0.192 to 0.821 of (M2), and the estimate increases from 0.190 to 0.794 of (M3). Although the estimate from measurement error mechanism (M1) is generally smaller than both that of (M2) and (M3), the variability of these estimates for each fixed $\beta$ is small. In the absence of the fever killing effect and measurement error, the estimates from other model approaches were about 0.177, however, the estimate of MAFF is inflated to about 0.190 after considering the measurement error problem (no fever killing means that there is only measurement error). Moreover, when the fever killing effect is severe such as 95\% killing effect, the estimates of the MAFF are about 0.8. Figure~\ref{fig:maff} visually shows the results of Table~\ref{tab:maff}. The figure is plotting the estimates of the MAFF vs. the values of $1-\beta$. As can be seen in Figure~\ref{fig:maff}, the effect of fever killing is moderate for a moderate size of fever killing, but a large killing effect boosts the estimates of the models (M1)-(M3). One distinct pattern between the three measurement error mechanism models is that the estimates of the models (M2) and (M3) are similarly affected as a fever killing effect increases and are similar for every value of $\beta$. 

\begin{figure}
\centerline{
\includegraphics[width=160mm]{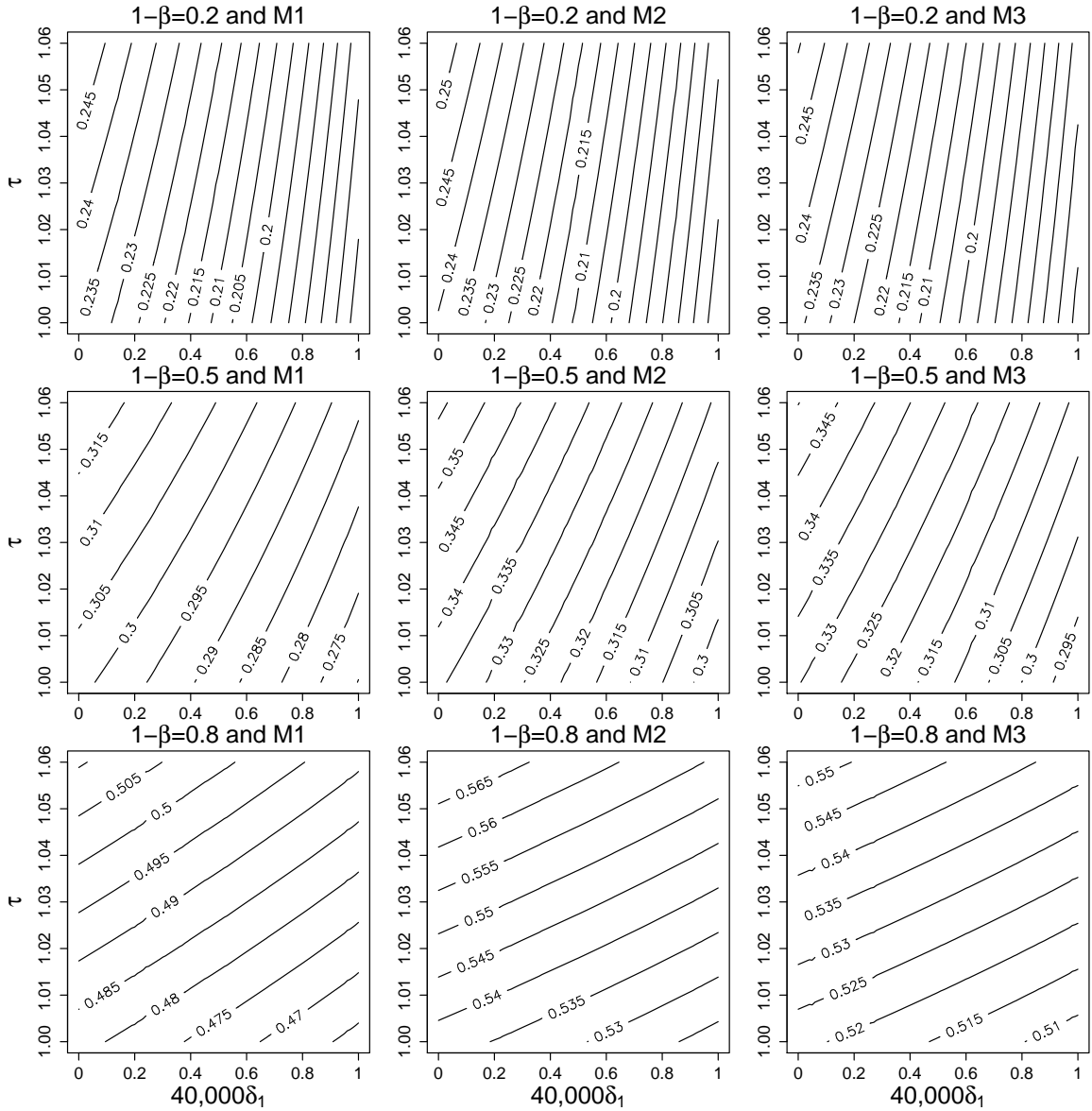}}
\caption{Sensitivity Analysis for violation of Assumption~\ref{assumption1}. From top to bottom, the sizes of fever killing are 20\%, 50\%, 80\%. From left to right, the measurement error mechanisms are M1, M2, M3. The estimates of the MAFF are represented as contour levels according to the values of $\delta_1$ and $\tau$. The scaled x-axis 40,000 $\times \delta_1$ is used rather than $\delta_1$ because of 0 $\leq 40,000 \times \delta_1 \leq 1$. The difference between adjacent contour levels is 0.005 in every sensitivity analysis.}
\label{fig:sensi}
\end{figure}

In the analysis of the Tanzania malaria data, we have demonstrated that the estimate of the MAFF not only depends on the fever killing size, but also depends on the measurement error mechanism. To obtain more precise results, it is required to know both a plausible range of the fever killing size and which type of measurement error occurs a priori.

We also conduct an additional sensitivity analysis for violation of Assumption~\ref{assumption1} (iii). We use two sensitivity parameters $\delta_1$ and $\tau$ discussed in Section~\ref{ss:violation}. The sensitivity parameter $\delta_1$ represents how two densities $f(\Dno=x | \Ymistar=0, \Ynmi=0)$ and $f(\Dno=x | \Ymistar=0, \Ynmi=1)$ differ ($\delta_1=0$ means that the densities are equal), and the sensitivity parameter $\tau$ represents the relative rate of not having a non-malarial fever according to $\Ymistar$. We consider plausible ranges of the parameters: $0 \leq \delta_1 \leq 1/40,000$ (this range is wide enough because the number of parasites is between 0 and 100,000) and $1 \leq \tau \leq 1.06$. Figure~\ref{fig:sensi} shows the  estimates of the MAFF according to the values of $\delta_1$ and $\tau$ for a size of fever killing $\beta$ ($\beta=0.2, 0.5, 0.8$) and a measurement error mechanism (M1, M2, M3). The estimates are plotted by using contour plots. The effect of deviation from Assumption~\ref{assumption1} (iii) is shown differently according to which factor caused a violation. As $\delta_1$ increases, the estimate of the MAFF decreases and as $\tau$ increases, the estimate increases. Therefore, a mixed effect of $\delta_1$ and $\tau$ appears to cancel each other out and to have a slight impact on the estimate. Another noticeable pattern is that the impact of $\delta_1$ is severer than the impact of $\tau$ when a fever killing effect is small, but this pattern is reversed when a fever killing effect is large. 

\section{SUMMARY}
\label{s:summary}

In this article, we have proposed a new approach to estimate the MAFF in the presence of both fever killing and measurement error. We have shown that existing MAFF estimators can be substantially biased in the presence of these problems. We develop a new estimator using the g-modeling approach to the Bayes deconvolution problem. To develop this new estimator, we extended the existing g-modeling approach that solves the convolution problem in non-mixture data to a setting of two-component mixture data such as malaria data. Under the assumptions that the size of fever killing effect is known and measurement error mechanism is known, our new estimator performs well. In practice, when the size of the fever killing effect is not known, we recommend choosing a plausible range of the fever killing size and comparing the corresponding estimates for a sensitivity analysis. To get a better estimate of the MAFF, further research is needed to better understand the fever killing effect to be able to get a narrower range of the fever killing size. Another difficulty in practice is to specify the measurement error mechanism. This problem can be eased by considering several plausible measurement error mechanism models from the simplest to the most complicated as we did in Section~\ref{s:example}. If a more complicated model produces a similar estimate to that of a less complicated model in general, then we can be more confident with our conclusion. More research on understanding the measurement error mechanism would also be useful.

\appendix
\section{APPENDIX}
\subsection{Proof of Proposition~\ref{proposition1}}
\label{appx:prop1}

Since $Y^{nmi}$ does not depend on the parasite level $D$ and $P(Y^{mi}=1 | D=0)=0$, we can have 
\begin{equation}
\begin{array}{ccl}
P(Y^{obs}=1 | D=0) & = & P(Y^{nmi}=1, Y^{mi}=0 | D=0) + P(Y^{mi}=1 | D=0) \\
&=& P(Y^{nmi}=1) P(Y^{mi}=0 | D=0)  \\
&=& P(Y^{nmi}=1)
\end{array}
\label{appx:eqn1}
\end{equation}

Similarly, we have 
\begin{equation}
\begin{array}{ccl}
P(Y^{obs}=1 | D>0) & = & P(Y^{nmi}=1 | D>0) + P(Y^{nmi}=0, Y^{mi}=1 | D>0) \\
&=& P(Y^{nmi}=1) + P(Y^{nmi}=0, Y^{mi}=1 | D>0)  \\
\end{array}
\label{appx:eqn2}
\end{equation}

Then, from Equations~(\ref{appx:eqn1}) and (\ref{appx:eqn2}), $\widehat{MAFF}_{RR}$ is 
\begin{equation}
\begin{array}{ccl}
\widehat{MAFF}_{RR} & = & p_f (R-1)/R \\
& = & P(D>0 | Y^{obs}=1) \cdot \frac{P(Y^{obs}=1 | D>0)-P(Y^{obs}=1 | D=0)}{P(Y^{obs}=1 | D>0)}\\
&=& P(D>0 | Y^{obs}=1) \cdot \frac{\{P(Y^{nmi}=1) + P(Y^{nmi}=0, Y^{mi}=1 | D>0)\} - P(Y^{nmi}=1)}{P(Y^{obs}=1 | D>0)} \\
&=& \frac{P(Y^{obs}=1, D>0)}{P(Y^{obs}=1)} \frac{P(Y^{nmi}=0, Y^{mi}=1 | D>0)}{P(Y^{obs}=1 | D>0)} \\
&=& \frac{P(Y^{nmi}=0, Y^{mi}=1, D>0)}{P(Y^{obs}=1)} \\
&=& P(Y^{nmi}=0, Y^{mi}=1 | Y^{obs}=1)
\end{array}
\label{appx:eqn3}
\end{equation}

\subsection{Proof of Proposition~\ref{proposition2}}
\label{appx:prop2}

Let $R^*$ be the odds ratio $p_f(1-p_a)/p_a(1-p_f)$. Since $p_f = P(D>0 | Y^{obs}=1)$ and $p_a = P(D>0 | Y^{obs}=0)$, the odds ratio $R^*$ is 
\begin{equation}
\begin{array}{ccl}
R^* &=& \frac{p_f}{1-p_f} \cdot \frac{1-p_a}{p_a} \\
&=& \frac{P(D>0 | Y^{obs}=1)}{P(D=0 | Y^{obs}=1)} \cdot \frac{P(D=0 | Y^{obs}=0)}{P(D>0 | Y^{obs}=0)} \\
&=& \frac{P(Y^{obs}=1, D>0)}{P(Y^{obs}=1, D=0)} \cdot \frac{P(Y^{obs}=0, D=0)}{P(Y^{obs}=0, D>0)} \\
&=& \frac{P(Y^{obs}=1, D>0)}{P(Y^{obs}=0, D>0)} \cdot \frac{P(Y^{obs}=0, D=0)}{P(Y^{obs}=1, D=0)} \\
&=& \frac{P(Y^{obs}=1 | D>0)}{P(Y^{obs}=0 | D>0)} \cdot \frac{P(Y^{obs}=0 | D=0)}{P(Y^{obs}=1 | D=0)}.
\end{array}
\label{appx:eqn4}
\end{equation}

By substituting Equation~(\ref{appx:eqn1}), $R^*$ is 
\begin{equation}
\begin{array}{ccl}
R^* &=& \frac{P(Y^{obs}=1 | D>0)}{P(Y^{nmi}=0, Y^{mi}=0 | D>0)} \cdot \frac{P(Y^{nmi}=0)}{P(Y^{nmi}=1)} \\
&=& \frac{P(Y^{obs}=1 | D>0)}{P(Y^{nmi}=1, Y^{mi}=0 | D>0)}.
\end{array}
\label{appx:eqn5}
\end{equation}

Then, $\widehat{MAFF}_{OR}$ is
\begin{equation}
\begin{array}{ccl}
\widehat{MAFF}_{OR} &=& p_f \cdot \frac{R^*-1}{R^*} \\
&=& P(D>0 | Y^{obs}=1) \cdot \frac{P(Y^{mi}=1 | D>0)}{P(Y^{obs}=1 | D>0)} \\
&=& \frac{P(Y^{mi}=1, D>0)}{P(Y^{obs}=1)} \\
&=& \frac{P(Y^{mi}=1)}{P(Y^{obs}=1)} 
\end{array}
\label{appx:eqn6}
\end{equation}

\subsection{Estimation of the dispersion parameter in the negative binomial distribution}
\label{appx:computation}

In Section 6, we assume that the measurement error model (M2) has the negative binomial distribution: $D^{obs}|D^{cur} \sim 40 \times NB(D^{cur}/40, r)$ where the mean is $D^{cur}/40$ and the dispersion parameter is $r$. The dispersion parameter $r$ is not known, but can be estimated from the data in \citet{o2007malaria}. 

\citet{o2007malaria} computed the false negative rate by counting numbers of slides reported as negative from 25 microscopists, and they plotted the false nagative rate on the mean parasite density in Figure 2. We let $y$ be the number of negative slides and $x$ be the mean parasite density. We use the data $(x, y)$ to find the maximum likelihood estimate of the dispersion parameter $r$; $y_i$ is the number of `negative' from the binomial distribution $B(n=25, p_i)$ where $p_i$ is the probability of being falsely negative, and $p_i$ is computed from the negative binomial distribution $NB(x_i/40, r)$. The log-likelihood is given as
\begin{equation}
\begin{array}{ccl}
\ell &\propto& \sum_{i=1}^{n} y_i \log(p_i) + (25-y_i) \log(1-p_i) \\
&=& \sum_{i=1}^{n} y_i \log(f(0; x_i/40, r)) + (25-y_i) \log(1-f(0; x_i/40, r)) 
\end{array}
\end{equation}
where $f(x; x_i/40, r)$ is the probability mass function of the negative binomial with the mean $x_i/40$ and the dispersion parameter $r$. From the data in \citet{o2007malaria}, the estimate of $r$ is obtained as 5.83, and we use $r=6$ in our paper. The R code for this estimation is provided online.

\bibliographystyle{agsm}
\bibliography{fke}
\end{document}